\newcommand{\Ab}{\mathbf{A}}
\newcommand{\Bb}{\mathbf{B}}
\newcommand{\Cb}{\mathbf{C}}

\newcommand{\Ib}{\mathbf{I}} %== identity

\newcommand{\Kb}{\mathbf{K}} %== dielectric permittivity

\newcommand{\Mb}{\mathbf{M}} %== characteristic tensor
\newcommand{\Nb}{\mathbf{N}}
\newcommand{\Tb}{\mathbf{T}}
\newcommand{\Wb}{\mathbf{W}} %== axial tensor of k

\newcommand{\eb}{\mathbf{e}} %== electric field =========

\newcommand{\gb}{\mathbf{g}}
\newcommand{\ub}{\mathbf{u}}
\newcommand{\vb}{\mathbf{v}}

\newcommand{\Kel}{\mathbb{K}} %== Kerr
\newcommand{\Mel}{\mathbb{M}} %== Maxwell tensor
\newcommand{\Nel}{\mathbb{N}} %== Neumann tensor

\newcommand{\Iel}{\mathbb{I}}

\newcommand{\bzero}{\mathbf{0}}

\newcommand{\Real}{\mathbb{R}}

\newcommand{\caV}{\mathcal V}

\newcommand{\caI}{\mathcal I}

\def\sym{\mathop{\rm sym}}
\def\Sym{\mathop{\rm Sym}}
\def\Sph{\mathop{\rm Sph}}
\def\Dev{\mathop{\rm Dev}}
\def\Lin{\mathop{\rm Lin}}

\def\tr{\mathop{\rm tr}}
\def\sph{\mathop{\rm sph}}
\def\dev{\mathop{\rm dev}}

\documentclass[preprint,12pt]{elsarticle}

\usepackage{amssymb}
\usepackage{epsfig}
\usepackage{epic}
\usepackage{pgfplots}
\usepackage{pgf}
\usepackage{tikz}
\pgfplotsset{compat=1.14}
\journal{ArXiv.org}

\begin{document}

\begin{frontmatter}



\title{Exact and linearized refractive index stress-dependence in anisotropic photoelastic crystals}
\author{Fabrizio Dav\'{\i}\footnote{e-mail: davi@univpm.it}}
\address{DICEA and ICRYS, Universit\'a Politecnica delle Marche\\
via Brecce Bianche, 60131 Ancona, Italy}
\begin{abstract}
For the permittivity tensor of photoelastic anisotropic crystals we obtain the exact non-linear dependence on the Cauchy stress tensor. We obtain the same result for its square root whose principal components, the crystal principal refractive index, are the starting point for any photoelastic analysis of transparent crystals. From these exact results then we obtain, in a total general manner, the linearized expressions to within higher-order terms in the stress tensor for both the permittivity tensor and its square root. We finish by showing some relavant examples of both non-linear and linearized relations for optically isotropic, uniaxial and biaxial crystals.
\end{abstract}

\begin{keyword}
Photoelastic crystals\sep Refractive index\sep Anisotropic crystals 
\MSC[2010] 74B99 \sep 74E10\sep 74F15\sep 78A10
\end{keyword}
\end{frontmatter}

This paper is dedicated to Paolo Podio-Guidugli for his $80^{th}$ birthday.

\section{Introduction}

In crystal photoelasticity the evaluation of the principal refractive index and their dependence on the stress, either applied or residual, is a mandatory step for any theoretical and experimental analysis of the optical properties of transparent crystals (\emph{vid. e.g.\/} \cite{BW99}-\cite{SS82}).

For $\Bb_{o}$ the dielectric permeablity tensor in the unstressed state and $\Mel$ the fourth-order piezo-optic tensor, the dielectric permeablity of a stressed crystal is described by the Maxwell linear relation:
\begin{equation}\label{inversepermittivity}
\Bb(\Tb)=\Bb_{o}+\Mel[\Tb]\,,
\end{equation}
where $\Tb$ is the symmetric Cauchy stress tensor. The relation between the principal values $(B_{1}\,,B_{2}\,,B_{3})$ of $\Bb(\Tb)$ and the principal refractive index $(n_{1}\,,n_{2}\,,n_{3})$ is:
\begin{equation}\label{principalK}
B_{k}=n_{k}^{-2}\,,\quad k=1,2,3\,;
\end{equation}
the difference between principal refractive index, the birifringence:
\begin{equation}\label{birifri}
\Delta n=n_{i}-n_{k}\,,\quad i,k=1,2,3\,,i\neq k\,,
\end{equation}
is one of the most important measurable quantities in photoelastic experiments (\emph{cf. e.g.\/} \cite{KR74}-\cite{BA19}).

Clearly, any analytical evaluation of the principal refractive index $n_{k}$ can be done provided that we are able first to write the inverse of $\Bb(\Tb)$, the dielectric permittivity
\begin{equation}\label{permittivity}
\Kb(\Tb)=\Bb^{-1}(\Tb)=(\Bb_{o}+\Mel[\Tb])^{-1}\,,
\end{equation}
and then to obtain its square root:
\begin{equation}\label{squareroot}
\Nb(\Tb)=\Kb^{\frac{1}{2}}(\Tb)\,,
\end{equation}
the principal values $(n_{1}\,,n_{2}\,,n_{3})$ of $\Nb(\Tb)$ being the principal refractive index.

The typical solution of this problem is to find first the eigencouples $(B_{k}\,,\ub_{k})$ of $\Kb(\Tb)$, then take the square root of the inverse of (\ref{principalK}) and finally, if we need linearized relations, linearize the result about the unstressed state $\Tb=\bzero$,  like we did for instance into \cite{DRM1} and \cite{DRM2}.  Such an approach has many limitations, since the possibility to find an explicit expression for the eigencouples $(B_{k}\,,\ub_{k})$ depends heavily on the crystal symmetry trough $\Mel$ and on the stress tensor $\Tb$: indeed into \cite{DRM1}, \cite{DRM2} we considered special state of stress. Moreover, for optically uniaxial materials the linearization about the unstressed state may be not well-defined since the derivative of $n_{k}$ with respect to $\Tb$ may blows-up for $\Tb\rightarrow\bzero$. 

Recently, searching for an easy way to represent the rotation in the polar decomposition of the deformation gradient, in a serendipitously way I found an old paper of Hoger and Carlson \cite{HC84} dealing with the inversion of a tensor like (\ref{inversepermittivity}) and with the square-root extraction like in (\ref{squareroot}). The most interesting thing is that the exact results obtained there didn't require an a-priori solution of an eigenvalue problem: rather they were obtained by a repeated application of the Cayley-Hamilton theorem.

In our paper we apply the results presented into \cite{HC84} to obtain explicit, exact and non-linear relations for the permittivity tensor (\ref{permittivity}) and for its square root (\ref{squareroot}), in terms of $\Bb_{o}$ and $\Mel[\Tb]$. Then, by starting from these exact results, we give a general linearization procedure which leads, to within higher-order terms into $\Tb$, to two relations which are equivalent to (\ref{inversepermittivity}), namely:
\begin{eqnarray}\label{finalrelations}
\Kb(\Tb)&=&\Kb_{o}+\Kel[\Tb]\,,\nonumber\\
&&\\
\Nb(\Tb)&=&\Nb_{o}+\Nel[\Tb]\,,\nonumber
\end{eqnarray}
with the two fourth-order piezo-optic tensor $\Kel$ and $\Nel$ expressed solely in terms of the components of the eigencouples of $\Bb_{o}$ and $\Mel$.

As a matter of fact however,  in order to obtain the principal refractive index from $\Nb(\Tb)$ we still need to solve an eigenvalues problem: besides special cases of stress in which one eigenvector of $\Nb(\Tb)$ is known, we have to solve the problem by the means of an approximate method like \emph{e.g.\/} that proposed into \cite{PT83}; on the other hand no further approximations besides the linearization, and no special hypothesis on $\Tb$ are necessary to obtain (\ref{finalrelations})$_{2}$.

\subsection*{Notation}

Let $\caV$ be the three-dimensional vector space whose elements we denote $\vb\in\caV$ and $\Lin$ the space of second order tensors $\Ab\in\Lin$. For $\{\eb_{k}\}\,,k=1,2,3$ an orthonormal base in $\caV$, the components of $\vb$ and $\Ab$ are given by $v_{i}=\vb\cdot\eb_{i}$ and $A_ {ij}=\Ab\cdot\eb_{i}\otimes\eb_{j}=\Ab\eb_{j}\cdot\eb_{i}$, $ i,j=1,2,3$.
We denote $\Sym$ and $\Sym^{+}$the subspaces of $\Lin$ of symmetric and positive-definite symmetric tensors respectively; in $\Sym$ we find useful to use the orthogonal base $\{\Wb_{h}\}\,,h=1,\ldots 6$:\footnote{We make use here of the Voigt's two index notation for second and fourth-order tensors, provided the identification $11=1\,,22=2\,,33=3$, $23=4\,,13=5\,,12=6$.}
\begin{equation}\label{baseinsym}
\begin{array}{ccc}
 \Wb_{1}=\eb_{1}\otimes\eb_{1}\,, & \Wb_{2}=\eb_{2}\otimes\eb_{2}\,,  &  \Wb_{3}=\eb_{3}\otimes\eb_{3}\,, \\
 \Wb_{4}=\sym(\eb_{2}\otimes\eb_{3})\,, & \Wb_{5}=\sym(\eb_{1}\otimes\eb_{3})\,,  &  \quad\Wb_{6}=\sym(\eb_{1}\otimes\eb_{2})\,, 
\end{array}
\end{equation}
with $\Ib=\Wb_{1}+\Wb_{2}+\Wb_{3}$. We define the spherical and deviatorical parts of $\Tb\in\Sym$ as:
\begin{equation}\label{sfericodev}
\sph\Tb=\sigma_{m}\Ib\,,\quad\dev\Tb=\Tb-\sph\Tb\,,\quad\sigma_{m}=\frac{1}{3}\tr\Tb\,,
\end{equation}
the underlying associated subspaces of $\Sym$ being $\Sph$ and $\Dev$, $\Sym=\Sph\oplus\Dev$; in the base (\ref{baseinsym}) we have:
\begin{equation}
\dev\Tb=\hat{T}_{11}\Wb_{1}+\hat{T}_{22}\Wb_{2}+\hat{T}_{33}\Wb_{3}+T_{23}\Wb_{4}+T_{13}\Wb_{5}+T_{12}\Wb_{6}\,,
\end{equation}
where
\begin{equation}
\hat{T}_{11}=\frac{2T_{11}-T_{22}-T_{33}}{3}\,,\quad\hat{T}_{22}=\frac{2T_{22}-T_{11}-T_{33}}{3}\,,\quad\hat{T}_{33}=\frac{2T_{33}-T_{11}-T_{22}}{3}\,.
\end{equation}
If $(\sigma_{k}\,,\eb_{k})$,  $k=,1,2,3$ are the eigencouples of $\Tb$ then by the decomposition (\ref{sfericodev}) we have
\begin{equation}\label{compoisotro}
\sigma_{m}=\frac{1}{3}(\sigma_{1}+\sigma_{2}+\sigma_{3})\,,\quad\dev\Tb=\hat{\sigma}_{k}\Wb_{k}\,,\quad\sigma_{k}-\sigma_{m}\,,\quad k=1,2,3\,,
\end{equation}
where $\hat{\sigma}_{k}$ are the eigenvalues of $\dev\Tb$.

The orthogonal invariants $\iota_{kA}$, $k=1,2,3$ of $\Ab\in\Lin$ are defined by:
\begin{eqnarray}
\iota_{1A}&=&\Ib\cdot\Ab=\tr\Ab\,,\nonumber\\
\iota_{2A}&=&\Ib\cdot\Ab^{*}=\frac{1}{2}((\tr\Ab)^{2}-\|\Ab\|^{2})\,,\\
\iota_{3A}&=&\det\Ab\,;\nonumber
\end{eqnarray}
here $\Ab^{*}=(\det\Ab)\Ab^{-T}$ denotes the cofactor of $\Ab$.  For $\alpha\in\Real$ the following identity holds:
\begin{equation}
\det(\Ab+\alpha\Ib)=\alpha^{3}+\alpha^{2}\iota_{1A}+\alpha\iota_{2A}+\iota_{3A}\,;
\end{equation}
moreover, for $\Cb=\Ab+\alpha\eb\otimes\eb+\beta\gb\otimes\gb$ with  $\alpha\,,\beta\in\Real$ and $\|\eb\|=\|\gb\|=1\,,\eb\cdot\gb=0$, it is:
\begin{eqnarray}\label{invariant2}
\iota_{1C}&=&\iota_{1A}+\alpha+\beta\,,\nonumber\\
\iota_{2C}&=&\iota_{2A}+\alpha\beta+\iota_{1A}(\alpha+\beta)-\Ab\cdot(\alpha\eb\otimes\eb+\beta\gb\otimes\gb)\,,\\
\iota_{3C}&=&\iota_{3A}+\Ab^{*}\cdot(\alpha\eb\otimes\eb+\beta\gb\otimes\gb)\,.\nonumber
\end{eqnarray}

Let $\Mel:\Sym\rightarrow\Sym$ be the piezo-optic fourth-order tensor, then its components are defined as
\begin{equation}
\Mel_{ijhk}=\Mel[\eb_{h}\otimes\eb_{k}]\cdot\eb_{i}\otimes\eb_{j}\,,\quad\Mel_{ijhk}=\Mel_{jihk}=\Mel_{ijkh}\,,
\end{equation}
or in the Voigt's two index notation:
\begin{equation}
\Mel_{AB}=\Mel[\Wb_{B}]\cdot\Wb_{A}\,,\quad A,B=1,\ldots 6\,.
\end{equation}
We denote $\Iel$ the fourth-order identity and with $\Mel^{T}$ the transpose of $\Mel$:
\begin{equation}
\Mel[\Ab]\cdot\Bb=\Mel^{T}[\Bb]\cdot\Ab\,,\quad\forall\Ab\,,\Bb\in\Lin\,.
\end{equation}
For given $\Bb\,,\Cb\in\Lin$ we shall make use of  the two fourth-order tensors $\Bb\boxtimes\Cb$ and $\Bb\otimes\Cb$ defined by:
\begin{equation}
(\Bb\boxtimes\Cb)\Mel[\Ab]=\Bb\Mel[\Ab]\Cb\,,\quad (\Bb\otimes\Cb)\Mel[\Ab]=(\Mel[\Ab]\cdot\Cb)\Bb\,,\quad\forall\Ab\,.
\end{equation}

\section{The dielectric permittivity tensor}

In \cite{HC84} it was obtained an analytical exact expression of the inverse:
\begin{equation}\label{inverse}
(c\Ib+\Mb)^{-1}\,,
\end{equation}
provided $c>0$ and $\Mb\in\Sym^{+}$; the result was obtained by the means of a repeated application of the Cayley-Hamilton theorem. Further, by the same tool, an explicit and exact analytical expression for the square root of a symmetric positive definite tensor was also given.

In the following subsections we shall show how the result given into \cite{HC84} allow for an explicit, exact and non-linear expression of the permittivity tensor $\Kb(\Tb)$ defined by (\ref{permittivity}) and of its square root $\Nb(\Tb)$ defined by (\ref{squareroot}).

We shall treat separately the three cases of Optically Isotropic, Optically Uniaxial and Optically Biaxial crystals, which differs for the different multiplicity of the eigenvalues of $\Bb_{o}$.

\subsection{Optically Isotropic crystals}

For Optically Isotropic crystals (which are comprised of Isotropic materials and Cubic crystals), the tensor $\Bb_{o}\in\Sph$, with
\begin{equation}\label{boisotro}
\Bb_{o}=n_{o}^{-2}\Ib\,.
\end{equation}

Hence, the results of \cite{HC84} can be used directly provided in (\ref{inverse}) we identify:
\begin{equation}
c=n_{o}^{-2}\,,\quad\Mb=\Mel[\Tb]\,.
\end{equation}

We notice that, whereas both $\Bb(\Tb)\in\Sym^{+}$ and $\Bb_{o}\in\Sym^{+}$, nothing can be said about the difference $\Bb(\Tb)-\Bb_{o}=\Mel[\Tb]$: however in \cite{HC84}, the positive-definiteness of $\Mb$ is an invertibility requirement; accordingly we simply made the weaker assumption that $\Mb=\Mel[\Tb]$ is invertible for all $\Tb$. 

Granted such an assumption, from equation (2.2) of \cite{HC84} we obtain the explicit representation for $\Kb(\Tb)$:
\begin{equation}\label{kappanonlinear}
\Kb(\Tb)=\frac{1}{\alpha_{3}}\bigg(\alpha_{1}\Ib-\alpha_{2}\Mel[\Tb]+\Mel[\Tb]^{2}\bigg)\,,
\end{equation}
where the three functions $\alpha_{j}\,,j=1,2,3$ are given by
\begin{eqnarray}\label{functions1}
\alpha_{1}(n_{o}\,,\Mel[\Tb])&=&n_{o}^{-4}+n_{o}^{-2}\iota_{1M}+\iota_{2M}\,,\nonumber\\
\alpha_{2}(n_{o}\,,\Mel[\Tb])&=&n_{o}^{-2}+\iota_{1M}\,,\\
\alpha_{3}(n_{o}\,,\Mel[\Tb])&=&n_{o}^{-6}+n_{o}^{-4}\iota_{1M}+n_{o}^{-2}\iota_{2M}+\iota_{3M}\,,\nonumber
\end{eqnarray}
and the orthogonal invariants of $\Mel[\Tb]$ by:
\begin{eqnarray}
\iota_{1M}&=&\tr\Mel[\Tb]\,,\nonumber\\
\iota_{2M}&=&\frac{1}{2}((\tr\Mel[\Tb])^{2}-\|\Mel[\Tb]\|^{2})\,,\\
\iota_{3M}&=&\det\Mel[\Tb]\neq 0\,.\nonumber
\end{eqnarray}

We remark that relation (\ref{kappanonlinear}) can also be arrived directly by the representation theorem for isotropic functions (\emph{vid. e.g.\/} \cite{GU81}):
\begin{equation}
\Kb(\Mb)=\caI_{o}(\iota_{kM})\Ib+\caI_{1}(\iota_{kM})\Mb+\caI_{2}(\iota_{kM})\Mb^{2}\,;
\end{equation}
in our treatment the depencence of the three functions $\caI_{j}(\iota_{kM})$, $j=0,1,2$ on the invariants of $\Mb$ is made explicitly by (\ref{functions1}) as a consequence of the procedure followed into \cite{HC84}.

To obtain the square root of $\Kb(\Tb)$ we then make use of formula (3.7) of \cite{HC84} which gives $\Nb(\Tb)$ in terms of functions of the invariants of both $\Kb(\Tb)$ and $\Nb(\Tb)$: 
\begin{equation}\label{enneexact}
\Nb(\Tb)=\frac{1}{\nu_{4}}\bigg(\nu_{1}\Ib+\nu_{2}\Kb(\Tb)-\nu_{3}\Kb(\Tb)^{2}\bigg)\,,
\end{equation}
where
\begin{eqnarray}\label{15}
\nu_{1}&=&\iota_{1N}\iota_{3K}+\iota_{3N}(\iota_{2N}(\iota_{2N}+\iota_{1K})+\iota_{2K})\,,\nonumber\\
\nu_{2}&=&(\iota_{1N}\iota_{2N}-\iota_{3N})(\iota_{2N}+\iota_{1K})\,,\\
\nu_{3}&=&\iota_{1N}\iota_{2N}-\iota_{3N}\,,\nonumber\\
\nu_{4}&=&\iota_{2N}(\iota_{2N}(\iota_{2N}+\iota_{1K})+\iota_{2K})+\iota_{3K}\,;\nonumber
\end{eqnarray}
the invariants $\iota_{kN}$ in (\ref{15}) can be represented explicitly in terms of the invariants $\iota_{kK}$ (\emph{vid.\/} \cite{HC84}, \S.5 and eqn.s (D2) and (D5) in the Dataset [DS] for this paper).

If we use (\ref{kappanonlinear}) into (\ref{enneexact}) we arrive at the explicit expression for the square root of the permittivity tensor:
\begin{equation}\label{Nfinal}
\Nb(\Tb)=a_{o}\Ib+a_{1}\Mel[\Tb]+a_{2}\Mel[\Tb]^{2}+a_{3}\Mel[\Tb]^{3}+a_{4}\Mel[\Tb]^{4}\,,
\end{equation}
where the five functions $a_{j}=a_{j}(n_{o}\,,\Mel[\Tb])=a_{j}(\alpha_{k}\,,\nu_{h})\,,j=1,\ldots 5$ are given explicitly by:
\begin{eqnarray}\label{constantA}
a_{o}&=&\frac{\nu_{1}\alpha_{3}^{2}+\nu_{2}\alpha_{1}\alpha_{3}-\nu_{3}\alpha_{1}^{2}}{\nu_{4}\alpha_{3}^{2}}\,,\nonumber\\
a_{1}&=&-\frac{\nu_{2}\alpha_{2}\alpha_{3}-2\nu_{3}\alpha_{1}\alpha_{2}}{\nu_{4}\alpha_{3}^{2}}\,,\nonumber\\
a_{2}&=&\frac{\nu_{2}\alpha_{3}-2\nu_{3}\alpha_{1}-\nu_{3}\alpha_{2}}{\nu_{4}\alpha_{3}^{2}}\,,\\
a_{3}&=&\frac{2\nu_{3}\alpha_{2}}{\nu_{4}\alpha_{3}^{2}}\,,\nonumber\\
a_{4}&=&-\frac{\nu_{3}}{\nu_{4}\alpha_{3}^{2}}\,.\nonumber
\end{eqnarray}
The principal values of (\ref{Nfinal}) then gives the explicit and non-linear formula for the principal refraction index. 

\subsection{Optically Uniaxial crystals}

Optically uniaxial crystals belongs to Trigonal, Tetragonal and Hexagonal symmerty groups. In all these cases the tensor $\Bb_{o}$ for uniaxial crystal has representation, provided we identify the optic axis direction with $\eb_{3}$:
\begin{equation}\label{bzerouniax}
\Bb_{o}=n_{o}^{-2}(\Ib-\Wb_{3})+n_{e}^{-2}\Wb_{3}=n_{o}^{-2}\Ib+(n_{e}^{-2}-n_{o}^{-2})\Wb_{3}\,,
\end{equation}
where $n_{o}$ is the ordinary and $n_{e}$ the extraordinary refractive index; accordingly
\begin{equation}
\Bb(\Tb)=n_{o}^{-2}\Ib+(n_{e}^{-2}-n_{o}^{-2})\Wb_{3}+\Mel[\Tb]\,,
\end{equation}
and the results obtained for optically isotropic materials still hold provided we set the tensor $\bar{\Mb}$ in place of $\Mb$:
\begin{equation}
\bar{\Mb}=\Mb+D\Wb_{3}\,,\quad\Mb=\Mel[\Tb]\,,\quad D=n_{e}^{-2}-n_{o}^{-2}\,.
\end{equation}

Equations (\ref{kappanonlinear}), (\ref{functions1}), (\ref{15}), (\ref{Nfinal}) and (\ref{constantA}) still hold provided we replace $\Mb$ with $\bar{\Mb}$, the relation between their invariants being obtained by the means of identities (\ref{invariant2}):
\begin{eqnarray}
\iota_{1\bar{M}}&=&\iota_{1M}+D\,,\nonumber\\
\iota_{2\bar{M}}&=&\iota_{2M}+D\Mb\cdot(\Ib-\Wb_{3})\,,\\
\iota_{3\bar{M}}&=&\iota_{3M}+D\Mb^{*}\cdot\Wb_{3}\,.\nonumber
\end{eqnarray}

The dielectric permittivity tensor for uniaxial crystal then is given by:
\begin{eqnarray}\label{kappanonlinearuni}
\Kb(\Tb)&=&\frac{1}{\beta_{3}}\bigg(\beta_{1}(\Ib-\Wb_{3})+(\beta_{1}-\beta_{2} D+D^{2})\Wb_{3}-\beta_{2}\Mel[\Tb]\\
&+&D(\Mel[\Tb]\Wb_{3}+\Wb_{3}\Mel[\Tb])+\Mel[\Tb]^{2}\bigg)\,,\nonumber
\end{eqnarray}
where the relation between the functions $\beta_{k}$, $k=1,2,3$ are given by: 
\begin{eqnarray}\label{functionbeta}
\beta_{1}(n_{o}\,,n_{e}\,,\Mb[\Tb])&=&\alpha_{1}+Dn_{o}^{-2}+D\Mb\cdot(\Ib-\Wb_{3})\,,\nonumber\\
\beta_{2}(n_{o}\,,n_{e}\,,\Mb[\Tb])&=&\alpha_{2}+D\,,\\
\beta_{3}(n_{o}\,,n_{e}\,,\Mb[\Tb])&=&\alpha_{3}+n_{o}^{-4}D+n_{o}^{-2}D\Mb\cdot(\Ib-\Wb_{3})+D\Mb^{*}\cdot\Wb_{3}\,.\nonumber
\end{eqnarray}

By setting (\ref{kappanonlinearuni}) into (\ref{enneexact}) then we get the non linear relation for $\Nb(\Tb)$ in uniaxial crystals: 
\begin{eqnarray}\label{enneuniaxial}
\Nb(\Tb)&=&b_{o}\Ib+b_{1}\Wb_{3}+b_{2}\Mb+b_{3}(\Mb\Wb_{3}+\Wb_{3}\Mb)+\nonumber\\
&+&b_{4}\Wb_{3}\Mb\Wb_{3}+b_{5}\Mb^{2}+b_{6}(\Mb\Wb_{3}\Mb+\Wb_{3}\Mb^{2}+\Mb^{2}\Wb_{3})\\
&+&b_{7}((\Mb\Wb_{3})^{2}+(\Wb_{3}\Mb)^{2}+\Wb_{3}\Mb^{2}\Wb_{3})+b_{8}\Mb^{3}\nonumber\\
&+&b_{9}(\Mb^{3}\Wb_{3}+\Wb_{3}\Mb^{3}+\Mb\Wb_{3}\Mb^{2}+\Mb^{2}\Wb_{3}\Mb)+b_{10}\Mb^{4}\,,\nonumber
\end{eqnarray}
where the eleven functions $b_{k}(n_{o}\,,n_{e}\,,\Mel[\Tb])=b_{k}(\beta_{j}\,,\nu_{h})\,,k=0,1,\ldots 10$ are given explicitly by eqn. (D8) of the dataset [DS].

\subsection{Optically Biaxial crystals}

Optically biaxial crystals are of the Triclinic, Monoclinic and Orthorhombic symmetry groups and have three different principal refractive index: if we assume that the orthonormal frame $\{\eb_{k}\}$, $k=1,2,3$ is also the principal frame for $\Bb_{o}$ then:\footnote{This is true for all biaxial crystals but Monoclinic and Triclinic. For Monoclinic crystals however the monoclinic $b$-axis is a principal direction and hence we can obtain an explicit representation for the eigencouples of $\Bb_{o}$.}
\begin{equation}\label{bzerobiax}
\Bb_{o}=B_{1}\Wb_{1}+B_{2}\Wb_{2}+B_{3}\Wb_{3}\,,\quad B_{k}=n_{k}^{-2}\,,k=1,2,3\,;
\end{equation}
let $B_{1}>B_{2}>B_{3}$, then we can rewrite (\ref{bzerobiax}) as:
\begin{equation}\label{bzerobiax1}
\Bb_{o}=B_{3}\Ib+D_{1}\Wb_{1}+D_{2}\Wb_{2}\,,\quad D_{\alpha}=B_{\alpha}-B_{3}\,,\alpha=1,2\,,
\end{equation}
and by replacing again into (\ref{kappanonlinear}), (\ref{functions1}), (\ref{enneexact}) and (\ref{15}) the tensor $\Mb$ with $\hat{\Mb}$
\begin{equation}
\hat{\Mb}=\Mb+D_{1}\Wb_{1}+D_{2}\Wb_{2}\,,
\end{equation}
we arrive at the following relation for the permittivity tensor
\begin{eqnarray}\label{kappanonlinearbiaxial}
\Kb(\Tb)&=&\frac{1}{\gamma_{3}}\bigg(\gamma_{1}\Ib+\sum_{\alpha=1}^{2}(D_{\alpha}^{2}-\gamma_{2} D_{\alpha}+)\Wb_{\alpha}-\gamma_{2}\Mel[\Tb]\\
&+&\sum_{\alpha=1}^{2}D_{\alpha}(\Mel[\Tb]\Wb_{\alpha}+\Wb_{\alpha}\Mel[\Tb])+\Mel[\Tb]^{2}\bigg)\,,\nonumber
\end{eqnarray}
where the functions $\gamma_{k}=\gamma_{k}(n_{j}\,,\Mel[\Tb])\,,j,k=1,2,3$ are defined by
\begin{eqnarray}
\gamma_{1}&=&\alpha_{1}+n_{3}^{-2}(D_{1}+D_{2})+D_{1}\Mb\cdot(\Ib-\Wb_{1})+D_{2}\Mb\cdot(\Ib-\Wb_{2})\,,\nonumber\\
\gamma_{2}&=&\alpha_{2}+D_{1}+D_{2}\,,\\
\gamma_{3}&=&\alpha_{3}+n_{3}^{-4}(D_{1}+D_{2})+n_{3}^{-2}D_{1}D_{2}\nonumber\\
&+&\sum_{\alpha=1}^{2}(n_{3}^{-2}D_{\alpha}\Mb\cdot(\Ib-\Wb_{\alpha})+D_{\alpha}\Mb^{*}\cdot\Wb_{\alpha})\,,\nonumber
\end{eqnarray}
and the orthogonal invariants of $\hat{\Mb}$ are:
\begin{eqnarray}
\iota_{1\hat{M}}&=&\iota_{1M}+D_{1}+D_{2}\,,\\
\iota_{2\hat{M}}&=&\iota_{2M}+D_{1}D_{2}+D_{1}\Mb\cdot(\Ib-\Wb_{1})+D_{2}\Mb\cdot(\Ib-\Wb_{2})\,,\nonumber\\
\iota_{3\hat{M}}&=&\iota_{3M}+\Mb^{*}\cdot(D_{1}\Wb_{1}+D_{2}\Wb_{2})\,.\nonumber
\end{eqnarray}

As in the previous cases by (\ref{enneexact}) and (\ref{kappanonlinearbiaxial}) we get the relation for the square root of the permittivity tensor for biaxial crystals:
\begin{eqnarray}\label{ennebiaxial}
\Nb(\Tb)&=&c_{o}\Ib+c_{1}\Wb_{1}+c_{2}\Wb_{2}+c_{3}\Mb+c_{4}(\Mb\Wb_{1}+\Wb_{1}\Mb)\nonumber\\
&+&c_{5}(\Mb\Wb_{2}+\Wb_{2}\Mb)+c_{6}\Wb_{1}\Mb\Wb_{1}\nonumber\\
&+&c_{7}\Wb_{2}\Mb\Wb_{2}+c_{8}(\Wb_{1}\Mb\Wb_{2}+\Wb_{2}\Mb\Wb_{1})\nonumber\\
&+&c_{9}\Mb^{2}+c_{10}\Mb\Wb_{1}\Mb+c_{11}\Mb\Wb_{2}\Mb\nonumber\\
&+&c_{12}((\Mb\Wb_{1})^{2}+(\Wb_{1}\Mb)^{2}+\Wb_{1}\Mb^{2}\Wb_{1})\\
&+&c_{13}((\Mb\Wb_{2})^{2}+\Wb_{2}\Mb)^{2}+\Wb_{2}\Mb^{2}\Wb_{2}))\nonumber\\
&+&c_{14}(\Wb_{1}\Mb^{2}\Wb_{2}+\Mb\Wb_{1}\Mb\Wb_{2}+\Wb_{1}\Mb\Wb_{2}\Mb)+c_{15}\Mb^{3}\nonumber\\
&+&c_{16}(\Mb\Wb_{1}\Mb^{2}+\Wb_{1}\Mb^{3}+\Mb^{3}\Wb_{1}+\Mb^{2}\Wb_{1}\Mb)\nonumber\\
&+&c_{17}(\Mb\Wb_{2}\Mb^{2}+\Wb_{2}\Mb^{3}+\Mb^{3}\Wb_{2}+\Mb^{2}\Wb_{2}\Mb)+c_{18}\Mb^{4}\,,\nonumber
\end{eqnarray}
with the nineteen functions $c_{k}=c_{k}(n_{j}\,,\Mel[\Tb])=c_{k}(\gamma_{j}\,,\nu_{h})\,,k=0,1,\ldots 18, j=1,2,3$ are given explicitly by eqn. (D10) of the dataset [DS].

\section{Linearized relations}

Relation (\ref{inversepermittivity}) is a linear relation in the stress $\Tb$: on the other hand, the exact relations for $\Kb(\Tb)$ and $\Nb(\Tb)$ we obtained in \S.2 are non-linear, involving the inversion of (\ref{inversepermittivity}) and the extraction of its square root. 

Crystals however are brittle materials, with a limited elastic range and a low brittle fracture tensile strength: accordingly it makes sense to consider an expression for the principal refractive index which is linearized in the stress, to arrive at a relation which is equivalent to (\ref{inversepermittivity}).

In previous papers dedicated to the same problem \cite{DRM1}, \cite{DRM2}, we obtained linear relations for $n_{k}(\Tb)$ by a linearization procedure which involved the eigenvalues $B_{k}$ of $\Bb(\Tb)$:
\begin{equation}
n_{k}(\Tb)=\frac{1}{\sqrt{B_{k}}}\bigg|_{\Tb=\bzero}-\frac{1}{2\sqrt{B_{k}}}\frac{\partial B_{k}}{\partial\Tb}\bigg|_{\Tb=\bzero}\cdot\Tb+o(\|\Tb\|^{2})\,;
\end{equation}
this procedure however was far from general (we must have explicit relations for $B_{k}$, which is not possible for all crystallographic classes and all stress) and moreover the derivative of $B_{k}$ blows-up to infinity for $\Tb=\bzero$ for uniaxial crystals and some special cases of stress.

We give here for both the permittivity tensor and for its square root a general linearization scheme which holds for any stress $\Tb$ and leads to (\ref{finalrelations})
\begin{eqnarray}\label{linearizedtwo}
\Kb(\Tb)&=&\Kb_{o}+\Kel[\Tb]+o(\|\Tb\|^{2})\,,\nonumber\\
&&\\
\Nb(\Tb)&=&\Nb_{o}+\Nel[\Tb]+o(\|\Tb\|^{2})\,,\nonumber
\end{eqnarray}
with $\Kb_{o}=\Kb(\bzero)$, $\Nb(\bzero)=\Nb_{o}$ and the two fourth-order tensors
\begin{equation}\label{KappaEnne}
\Kel=\frac{\partial\Kb}{\partial\Tb}\Bigg|_{\Tb=\bzero}\,,\quad\Nel=\frac{\partial\Nb}{\partial\Tb}\Bigg|_{\Tb=\bzero}\,;
\end{equation}
as in the previous section we shall treat in order the three cases of Optically Isotropic, Uniaxial and Biaxial crystals.

\subsection{Optically Isotropic}

In this case, since $\Bb_{o}=n_{o}^{-2}\Ib$, it is trivial to evaluate its inverse and the associated square root.
\begin{equation}\label{directiso}
\Kb(\bzero)=n_{o}^{2}\Ib\,,\quad\Nb(\bzero)=n_{o}\Ib\,.
\end{equation}
As far as the fourth-order tensor $\Kel$ is concerned, from (\ref{kappanonlinear}) we have
\begin{equation}
\Kel=\Ib\otimes\frac{\partial}{\partial\Tb}(\frac{\alpha_{1}}{\alpha_{3}})\Bigg|_{\Tb=\bzero}-\frac{\alpha_{2o}}{\alpha_{3o}}\,\Mel\,,
\end{equation}
where the terms $\alpha_{jo}=\alpha_{j}(n_{o}\,,\bzero)\,,j=1,2,3$ are 
\begin{equation}
\alpha_{1o}=n_{o}^{-4}\,,\quad\alpha_{2o}=n_{o}^{-2}\,,\quad\alpha_{3o}=n_{o}^{-6}\,;
\end{equation}
then, since
\begin{eqnarray}\label{derivatazeroiso}
\frac{\partial}{\partial\Tb}(\frac{\alpha_{1}}{\alpha_{3}})\Bigg|_{\Tb=\bzero}&=&\frac{1}{\alpha_{3o}^{2}}(\alpha_{3o}\frac{\partial\alpha_{1}}{\partial\Tb}\Bigg|_{\Tb=\bzero}-\alpha_{1o}\frac{\partial\alpha_{3}}{\partial\Tb}\Bigg|_{\Tb=\bzero})\\
&=&\frac{1}{\alpha_{3o}^{2}}(\alpha_{3o}n_{o}^{-2}-\alpha_{1o}n_{o}^{-4})\Mel^{T}[\Ib]=\bzero\,;\nonumber
\end{eqnarray}
we have, to within higher-order terms:
\begin{equation}
\Kb(\Tb)=n_{o}^{2}\Ib-n_{o}^{4}\Mel[\Tb]\,.
\end{equation}

When we turn our attention to (\ref{linearizedtwo})$_{2}$, by (\ref{Nfinal}) we have that:
\begin{equation}
\Nel=\Ib\otimes\frac{\partial a_{o}}{\partial\Tb}\Bigg|_{\Tb=\bzero}+a_{1}(n_{o}\,,\bzero)\Mel\,,
\end{equation}
and since by eqn.s (D12) and (D16) from the dataset [DS] it is
\begin{equation}
\frac{\partial a_{o}}{\partial\Tb}\Bigg|_{\Tb=\bzero}=\bzero\,,\quad a_{1}(n_{o}\,,\bzero)=-\frac{n_{o}^{3}}{2}\,,
\end{equation}
then we are led, to within higher-order terms in the stress tensor, to
\begin{equation}\label{Nelisotro}
\Nb(\Tb)=n_{o}\Ib-\frac{n_{o}^{3}}{2}\Mel[\Tb]\,.
\end{equation}

The piezo-optic tensors $\Kel$ and $\Nel$ for an optically isotropic material accordingly admit the following simple representation:
\begin{equation}
\Kel=-n_{o}^{4}\Mel\,,\quad\Nel=-\frac{n_{o}^{3}}{2}\Mel\,,
\end{equation}
being both proportional to the Maxwell piezo-optic tensor $\Mel$.

\subsection{Optically Uniaxial}

For optically uniaxial crystals, from (\ref{bzerouniax}), trivially:
\begin{equation}\label{directuni}
\Kb(\bzero)=n_{o}^{2}(\Ib-\Wb_{3})+n_{e}^{2}\Wb_{3}\,,\quad\Nb(\bzero)=n_{o}(\Ib-\Wb_{3})+n_{e}\Wb_{3}\,,
\end{equation}
whereas from (\ref{kappanonlinearbiaxial}) we get:
\begin{eqnarray}
\frac{\partial\Kb}{\partial\Tb}\Bigg|_{\Tb=\bzero}&=&\Ib\otimes\frac{\partial}{\partial\Tb}(\frac{\beta_{1}}{\beta_{3}})\Bigg|_{\Tb=\bzero}+\Wb_{3}\otimes\frac{\partial}{\partial\Tb}(\frac{D(D-\beta_{2})}{\beta_{3}})\Bigg|_{\Tb=\bzero}\\
&-&\frac{\beta_{2o}}{\beta_{3o}}\,\Mel+\frac{D}{\beta_{3o}}(\Ib\boxtimes\Wb_{3}+\Wb_{3}\boxtimes\Ib)\Mel\,,\nonumber
\end{eqnarray}
where
\begin{eqnarray}\label{beta0}
\beta_{1o}&=&\beta_{1}(n_{o}\,,n_{e}\,,\bzero)=\alpha_{1o}+Dn_{o}^{-2}\,,\nonumber\\
\beta_{2o}&=&\beta_{2}(n_{o}\,,n_{e}\,,\bzero)=\alpha_{2o}+D\,,\\
\beta_{3o}&=&\beta_{3}(n_{o}\,,n_{e}\,,\bzero)=\alpha_{3o}+Dn_{o}^{-4}\,.\nonumber
\end{eqnarray}
By (\ref{functionbeta}) and (\ref{beta0}) it is:
\begin{equation}\label{derivatainzerouni}
\frac{\partial}{\partial\Tb}(\frac{\beta_{1}}{\beta_{3}})\Bigg|_{\Tb=\bzero}=\frac{1}{\beta_{3o}^{2}}(\beta_{3o}\frac{\partial\beta_{1}}{\partial\Tb}\Bigg|_{\Tb=\bzero}-\beta_{1o}\frac{\partial\beta_{3}}{\partial{\Tb}}\Bigg|_{\Tb=\bzero})\,;
\end{equation}
then, since:
\begin{eqnarray}\label{doppiederive}
\frac{\partial\beta_{1}}{\partial\Tb}\Bigg|_{\Tb=\bzero}&=&\frac{\partial\alpha_{1}}{\partial\Tb}\Bigg|_{\Tb=\bzero}+D\Mel^{T}[\Ib-\Wb_{3}]\,,\nonumber\\
&&\\
\frac{\partial\beta_{3}}{\partial\Tb}\Bigg|_{\Tb=\bzero}&=&\frac{\partial\alpha_{3}}{\partial\Tb}\Bigg|_{\Tb=\bzero}+Dn_{o}^{-2}\Mel^{T}[\Ib-\Wb_{3}]\,,\nonumber
\end{eqnarray}
by (\ref{beta0})$_{1,3}$, (\ref{doppiederive}) and (\ref{derivatazeroiso}), then from (\ref{derivatainzerouni}) we arrive at
\begin{equation}\label{unifinal1}
\frac{\partial}{\partial\Tb}(\frac{\beta_{1}}{\beta_{3}})\Bigg|_{\Tb=\bzero}=\bzero\,.
\end{equation}

In a similar manner:
\begin{equation}\label{derivatainzerouni2}
\quad\frac{\partial}{\partial\Tb}(\frac{D(D-\beta_{2})}{\beta_{3}})\Bigg|_{\Tb=\bzero}=-\frac{D}{\beta_{3o}^{2}}(\beta_{3o}\frac{\partial\beta_{2}}{\partial\Tb}\Bigg|_{\Tb=\bzero}+(D-\beta_{2o})\frac{\partial\beta_{3}}{\partial{\Tb}}\Bigg|_{\Tb=\bzero})\,;
\end{equation}
since by (\ref{beta0})$_{2}$
\begin{equation}
\frac{\partial\beta_{2}}{\partial\Tb}\Bigg|_{\Tb=\bzero}=\frac{\partial\alpha_{2}}{\partial\Tb}\Bigg|_{\Tb=\bzero}=\Mel^{T}[\Ib]\,,
\end{equation}
then by (\ref{derivatainzerouni2}), (\ref{beta0})$_{3}$ and (\ref{derivatazeroiso}) we arrive at
\begin{equation}\label{unifinal2}
\quad\frac{\partial}{\partial\Tb}(\frac{D(D-\beta_{2})}{\beta_{3}})\Bigg|_{\Tb=\bzero}=n^{2}_{o}(n_{o}^{2}-n_{e}^{2})^{2}\Mel^{T}[\Wb_{3}]\,.
\end{equation}

From (\ref{unifinal1}) and (\ref{unifinal2}) then we get, to within higher-order terms, the linearized relation for the permittivity:
\begin{eqnarray}
\Kb(\Tb)&=&n_{o}^{2}(\Ib-\Wb_{3})+n_{e}^{2}\Wb_{3}-n_{o}^{4}\Mel[\Tb]\nonumber\\
&+&n^{2}_{o}(n_{e}^{2}-n_{o}^{2})(\Mel[\Tb]\cdot\Wb_{3}))\Wb_{3}\\
&+&n_{o}^{2}(n_{e}^{2}-n_{o}^{2})(\Mel[\Tb]\Wb_{3}+\Wb_{3}\Mel[\Tb])\,.\nonumber
\end{eqnarray}

As far as the tensor $\Nb(\Tb)$ is concerned,, from (\ref{enneuniaxial}) we have 
\begin{eqnarray}
\frac{\partial\Nb}{\partial\Tb}\Bigg|_{\Tb=\bzero}&=&\Ib\otimes\frac{\partial b_{o}}{\partial\Tb}\Bigg|_{\Tb=\bzero}+\Wb_{3}\otimes\frac{\partial b_{1}}{\partial\Tb}\Bigg|_{\Tb=\bzero}\nonumber\\
&+&b_{2}(n_{o}\,,n_{e}\,,\bzero)\Mel\nonumber\\
&+&b_{3}(n_{o}\,,n_{e}\,,\bzero)(\Ib\boxtimes\Wb_{3}+\Wb_{3}\boxtimes\Ib)\Mel\\
&+&b_{4}(n_{o}\,,n_{e}\,,\bzero)(\Wb_{3}\boxtimes\Wb_{3})\Mel\,,\nonumber
\end{eqnarray}
and by the means of eqn.(D24) of [DS], we obtain the linearized expression for the square root of the permittivity tensor:
\begin{eqnarray}
\Nb(\Tb)&=&n_{o}(\Ib-\Wb_{3})+n_{e}\Wb_{3}-\frac{n_{o}^{3}}{2}\Bigg(\Mel[\Tb]+F_{1}(\xi)(\Mel[\Tb]\Wb_{3}+\Wb_{3}\Mel[\Tb])\nonumber\\
&+&F_{2}(\xi)\Wb_{3}\Mel[\Tb]\Wb_{3}\Bigg)\,,
\end{eqnarray}
where the functions $F_{k}=F_{k}(\xi)\,,\xi=n_{e}/n_{o}\,,k=1,2,$ are defined by (\emph{vid.\/}[DS]):
\begin{eqnarray}
F_{1}(\xi)&=&\frac{(1-\xi^{2})(\xi^{2}+2\xi+2)}{2(1+\xi)^{2}}\,,\\
F_{2}(\xi)&=&\frac{(1-\xi)^{2}}{(1+\xi)^{2}}\,,\nonumber
\end{eqnarray}
with $F_{k}(1)=0$.

The piezo-optic tensors $\Kel$ and $\Nel$ for optically uniaxial crystals have the following representation:
\begin{eqnarray}\label{linearizzatouniax}
\Kel&=&-n_{o}^{4}\Big[\Iel-(\xi^{2}-1)(\Wb_{3}\otimes\Wb_{3}+\Ib\boxtimes\Wb_{3}+\Wb_{3}\boxtimes\Ib)\Big]\Mel\,,\nonumber\\
&&\\
\Nel&=&-\frac{n_{o}^{3}}{2}\Big[\Iel+F_{1}(\xi)(\Ib\boxtimes\Wb_{3}+\Wb_{3}\boxtimes\Ib)+F_{2}(\xi)\Wb_{3}\boxtimes\Wb_{3}\Big]\Mel\,.\nonumber
\end{eqnarray}

\subsection{Optically Biaxial}

Since $\Bb_{o}$ for optically biaxial crystals is given by (\ref{bzerobiax}), then $\Kb(\bzero)$ and $\Nb(\bzero)$ admit the explicit representations:
\begin{equation}\label{directbia}
\Kb(\bzero)=n_{1}^{2}\Wb_{1}+n_{2}^{2}\Wb_{2}+n_{3}^{2}\Wb_{3}\,,\quad\Nb(\bzero)=n_{1}\Wb_{1}+n_{2}\Wb_{2}+n_{3}\Wb_{3}\,;
\end{equation}
the fourth-order derivative of permittivity with respect to $\Tb$ is
\begin{eqnarray}
\frac{\partial\Kb}{\partial\Tb}\Bigg|_{\Tb=\bzero}&=&\Ib\otimes\frac{\partial}{\partial\Tb}(\frac{\gamma_{1}}{\gamma_{3}})\Bigg|_{\Tb=\bzero}+\sum_{\alpha=1}^{2}\Wb_{\alpha}\otimes\frac{\partial}{\partial\Tb}(\frac{D_{\alpha}(D_{\alpha}-\gamma_{2})}{\gamma_{3}})\Bigg|_{\Tb=\bzero}\\
&-&\frac{\gamma_{2o}}{\gamma_{3o}}\,\Mel+\sum_{\alpha=1}^{2}\frac{D_{\alpha}}{\gamma_{3o}}(\Ib\boxtimes\Wb_{\alpha}+\Wb_{\alpha}\boxtimes\Ib)\Mel\,.\nonumber
\end{eqnarray}
Since, by a repeated application of the same procedure we used for uniaxial materials, we have:
\begin{eqnarray}
\frac{\partial}{\partial\Tb}(\frac{\gamma_{1}}{\gamma_{3}})\Bigg|_{\Tb=\bzero}&=&\bzero\,,\\
\quad\frac{\partial}{\partial\Tb}(\frac{D_{\alpha}(D_{\alpha}-\gamma_{2})}{\gamma_{3}})\Bigg|_{\Tb=\bzero}&=&n_{3}^{2}(n_{\alpha}^{2}-n_{3}^{2})\Mel^{T}[\Wb_{\alpha}]\,,\quad\alpha=1,2\,,\nonumber
\end{eqnarray}
then we obtain, to within higher-order terms, the linearized relation for the permittivity:
\begin{eqnarray}\label{kelbiaxial}
\Kb(\Tb)&=&n_{1}^{2}\Wb_{1}+n_{2}^{2}\Wb_{2}+n_{3}^{2}\Wb_{3}-n_{3}^{4}\Mel[\Tb]\nonumber\\
&+&\sum_{\alpha=1}^{2}n_{3}^{2}(n_{\alpha}^{2}-n_{3}^{2})(\Mel[\Tb]\cdot\Wb_{\alpha})\Wb_{\alpha}\\
&+&\sum_{\alpha=1}^{2}n_{3}^{2}(n_{\alpha}^{2}-n_{3}^{2})(\Mel[\Tb]\Wb_{\alpha}+\Wb_{\alpha}\Mel[\Tb])\,.\nonumber
\end{eqnarray}

We turn our attention to (\ref{KappaEnne})$_{2}$ and then, by (\ref{ennebiaxial}), (\ref{directbia})$_{2}$ and (\ref{linearizedtwo})$_{2}$ we get
\begin{eqnarray}\label{nelbiaxial}
\Nb(\Tb)&=&n_{1}\Wb_{1}+n_{2}\Wb_{2}+n_{3}\Wb_{3}+\Ib\otimes\frac{\partial c_{o}}{\partial\Tb}\Big|_{\Tb=\bzero}+\sum_{\alpha=1}^{2}\Wb_{\alpha}\otimes\frac{\partial c_{\alpha}}{\partial\Tb}\Bigg|_{\Tb=\bzero}\nonumber\\
&+&c_{3}(n_{k}\,,\bzero)\Mel[\Tb]+c_{4}(n_{k}\,,\bzero)(\Mel[\Tb]\Wb_{1}+\Wb_{1}\Mel[\Tb])\\
&+&c_{5}(n_{k}\,,\bzero)(\Mel[\Tb]\Wb_{2}+\Wb_{2}\Mel[\Tb])+c_{6}(n_{k}\,,\bzero)\Wb_{1}\Mel[\Tb]\Wb_{1}\nonumber\\
&+&c_{7}(n_{k}\,,\bzero)\Wb_{2}\Mel[\Tb]\Wb_{2}+c_{8}(n_{k}\,,\bzero)(\Wb_{1}\Mel[\Tb]\Wb_{2}+\Wb_{2}\Mel[\Tb]\Wb_{1})\,;\nonumber
\end{eqnarray}
since (\emph{vid.\/} [DS]):
\begin{equation}
\frac{\partial c_{o}}{\partial\Tb}\Bigg|_{\Tb=\bzero}=\bzero\,,\quad\frac{\partial c_{\alpha}}{\partial\Tb}\Bigg|_{\Tb=\bzero}=\bzero\,,\quad\alpha=1,2,
\end{equation}
and
\begin{equation}
c_{k}(n_{j}\,,\bzero)=n_{3}^{3}G_{k}(\xi_{1}\,,\xi_{2})\,,\quad k=3,4,\ldots 8\,,
\end{equation}
where the six functions $G_{k}\,,k=3,\ldots 8$ of the two parameters $\xi_{\alpha}=n_{\alpha}/n_{3}\,,\alpha=1,2$, given explicitly in [DS],  are such that $G_{3}(1\,,1)=1/2$ and $G_{j}(1\,,1)=0\,,j\neq 3$, then from (\ref{nelbiaxial}) we get
\begin{eqnarray}\label{nelbiaxialfinal}
\Nb(\Tb)&=&n_{1}\Wb_{1}+n_{2}\Wb_{2}+n_{3}\Wb_{3}-n_{3}^{3}\Big(G_{3}(\xi_{1}\,,\xi_{2})\Mel[\Tb]\nonumber\\
&+&G_{4}(\xi_{1}\,,\xi_{2})(\Wb_{1}\Mel[\Tb]+\Mel[\Tb]\Wb_{1})+G_{5}(\xi_{1}\,,\xi_{2})(\Wb_{2}\Mel[\Tb]+\Mel[\Tb]\Wb_{2})\nonumber\\
&+&G_{6}(\xi_{1}\,,\xi_{2})\Wb_{1}\Mel[\Tb]\Wb_{1}+G_{7}(\xi_{1}\,,\xi_{2})\Wb_{2}\Mel[\Tb]\Wb_{2}\\
&+&G_{8}(\xi_{1}\,,\xi_{2})(\Wb_{1}\Mel[\Tb]\Wb_{2}+\Wb_{2}\Mel[\Tb]\Wb_{2})\Big)\,;\nonumber
\end{eqnarray}

By (\ref{kelbiaxial}) and (\ref{nelbiaxialfinal}) then the tensors $\Kel$ and $\Nel$ for a biaxial crystal have the explicit form:
\begin{eqnarray}\label{doublebiaxial}
\Kel&=&-\Big[n_{3}^{4}\,\Iel-\sum_{\alpha=1}^{2}n_{3}^{2}(n_{\alpha}^{2}-n_{3}^{2})(\Wb_{\alpha}\otimes\Wb_{\alpha}+\Ib\boxtimes\Wb_{\alpha}+\Wb_{\alpha}\boxtimes\Ib)\Big]\Mel\,,\nonumber\\
&&\\
\Nel&=&-n_{3}^{3}\Big[G_{3}(\xi_{1}\,,\xi_{2})\Iel+G_{4}(\xi_{1}\,,\xi_{2})(\Ib\boxtimes\Wb_{1}+\Wb_{1}\boxtimes\Ib)\nonumber\\
&+&G_{5}(\xi_{1}\,,\xi_{2})(\Ib\boxtimes\Wb_{2}+\Wb_{2}\boxtimes\Ib)\nonumber\\
&+&G_{6}(\xi_{1}\,,\xi_{2})\Wb_{1}\boxtimes\Wb_{1}+G_{7}(\xi_{1}\,,\xi_{2})\Wb_{2}\boxtimes\Wb_{2}\nonumber\\
&+&G_{8}(\xi_{1}\,,\xi_{2})(\Wb_{1}\boxtimes\Wb_{2}+\Wb_{2}\boxtimes\Wb_{1})\Big]\Mel\,.\nonumber
\end{eqnarray}

\section{Examples}

\subsection{Optically Isotropic crystals}

\subsubsection{Isotropic materials}

For Isotropic materials the piezo-optic tensor has the representation \cite{AU03}
\begin{equation}\label{isotropicM}
\Mel=M_{1}\frac{1}{3}\Ib\otimes\Ib+M_{2}(\Iel-\frac{1}{3}\Ib\otimes\Ib)
\end{equation}
where the two piezo-optic moduli $M_{1}$ and $M_{2}$ are expressed in terms of the components $\Mel_{ijhk}=\Mel_{jihk}=\Mel_{ijkh}$ of the piezo-optic tensor by
\begin{equation}
M_{1}=\Mel_{1111}+2\Mel_{1122}\,,\quad M_{2}=\Mel_{1111}-\Mel_{1122}\,.
\end{equation}
Accordingly, by (\ref{isotropicM}) and (\ref{sfericodev}) we have
\begin{equation}
\Mel[\Tb]=M_{1}\sigma_{m}\Ib+M_{2}\dev\Tb\,.
\end{equation}
Then, since $\Bb_{o}$ is given by (\ref{boisotro}), then the permittivity tensor is then given by relation (\ref{kappanonlinear}) together with (\ref{isotropicM}):
\begin{equation}\label{Kiso}
\Kb(\Tb)=\frac{1}{\alpha_{3}}\bigg((\alpha_{1}-\alpha_{2} M_{1}\sigma_{m}+M_{1}^{2}\sigma_{m}^{2})\Ib+M_{2}(2M_{1}\sigma_{m}-\alpha_{2})\dev\Tb+M_{2}^{2}(\dev\Tb)^{2}\bigg)\,,
\end{equation}
where, explicitly
\begin{eqnarray}\label{functions2}
\alpha_{1}&=&n_{o}^{-2}\alpha_{2}+3M_{1}^{2}\sigma_{m}^{2}+M_{2}^{2}\|\dev\Tb\|^{2}\,,\nonumber\\
\alpha_{2}&=&n_{o}^{-2}+3M_{1}\sigma_{m}\,,\\
\alpha_{3}&=&n_{o}^{-2}\alpha_{1}+(M_{1}\sigma_{m})^{3}+M_{1}M_{2}^{2}\sigma_{m}\hat{\iota}_{2T}+M_{2}^{3}\hat{\iota}_{3T}\,,\nonumber
\end{eqnarray}
and with $\hat{\iota}_{2T}$ and $\hat{\iota}_{3T}$ the invariants of $\dev\Tb$. We write (\ref{Kiso}) in terms of (\ref{compoisotro}) together with (\ref{functions2}) and since (\ref{Kiso}) is still in spectral form, then we get directly the principal refractive index as the square root of the principal values of (\ref{Kiso}):
\begin{equation}\label{nkiso}
n_{k}=\sqrt{\frac{A_{k}}{B}}\,,\quad k=1,2,3\,,
\end{equation}
with
\begin{eqnarray}
A_{k}&=&n_{o}^{-4}+2n_{o}^{-2}M_{1}\sigma_{m}+M_{1}^{2}\sigma_{m}^{2}+M_{2}^{2}(\hat{\sigma}_{1}^{2}+\hat{\sigma}_{2}^{2}+\hat{\sigma}_{3}^{2})\nonumber\\
&-&\hat{\sigma}_{k}(M_{1}M_{2}\sigma_{m}-n_{o}^{-2}M_{2})+M_{2}^{2}\hat{\sigma}^{2}_{k}\,,\nonumber\\
&&\\
B&=&n_{o}^{-2}(n_{o}^{-2}(n_{o}^{-2}+3M_{1}\sigma_{m})+3M_{1}^{2}\sigma_{m}^{2}+M_{2}^{2}(\hat{\sigma}_{1}^{2}+\hat{\sigma}_{2}^{2}+\hat{\sigma}_{3}^{2}))\nonumber\\
&+&M_{1}^{3}\sigma_{m}^{3}+M_{1}M_{2}\sigma_{m}(\hat{\sigma}_{1}\hat{\sigma}_{2}+\hat{\sigma}_{1}\hat{\sigma}_{3}+\hat{\sigma}_{2}\hat{\sigma}_{3})+M_{2}^{3}\hat{\sigma}_{1}\hat{\sigma}_{2}\hat{\sigma}_{3}\,,\nonumber
\end{eqnarray}

When we consider the linearized relation (\ref{Nelisotro}) then we arrive directly at the very well-known formula:
\begin{equation}\label{linearisotropic}
n_{k}=n_{o}-\frac{n_{o}^{3}}{2}(M_{1}\sigma_{m}+M_{2}\hat{\sigma_{k}})\,,\quad k=1,2,3\,,
\end{equation}
which shows how, in order to have birifringence, an isotropic material must be loaded by a deviatoric stress. From (\ref{birifri}) and (\ref{linearisotropic}) then it follows trivially the well-known Brewster's law \cite{BRE}, \cite{DA19}:
\begin{equation}
\Delta n=-\frac{n_{o}^{3}}{2}M_{2}(\hat{\sigma_{i}}-\hat{\sigma_{k}})=-\frac{n_{o}^{3}}{2}M_{2}(\sigma_{i}-\sigma_{k})\,,\quad i,k=1,2,3\,, i\neq k\,.
\end{equation}

In order to compare the linearized relation (\ref{linearisotropic}) with the non-linear relation (\ref{nkiso}) we consider separately a spherical stress $\sigma_{m}$ and a generic deviatorical stress $\hat{\sigma}_{k}$ for a LG-812 Nd-doped glass \cite{WA80} with (at $\lambda=632$ nm):
\begin{equation}
n_{o}=1.49\,,\quad M_{1}=2,01\cdot 10^{-6}\mbox{mm}^{2}/\mbox{N}\,,\quad M_{2}=-0.63\cdot 10^{-6}\mbox{mm}^{2}/\mbox{N}\,;
\end{equation}
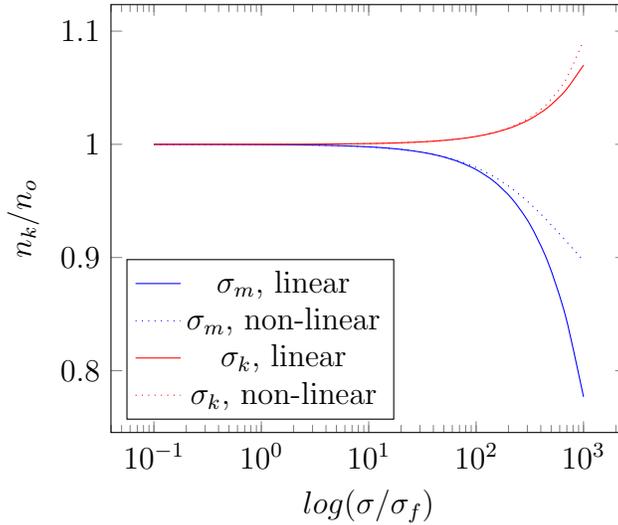
\begin{tikzpicture}
    \begin{semilogxaxis}[legend pos=south west,
        xlabel=$log(\sigma/\sigma_{f})$,
        ylabel=$n_{k}/n_{o}$
        ]

    \addplot[smooth, domain=0.1:1000, blue] {1-0.00022311*x};
    \addlegendentry{$\sigma_{m}$, linear}
    \addplot[smooth, domain=0.1:1000, dotted,blue] {0.6675*((0.202+0.000181*x)/(0.09+0.000122*x))^(0.5)};
    \addlegendentry{$\sigma_{m}$, non-linear}
    \addplot[smooth, domain=0.1:1000, red] {1+0.00007*x};
     \addlegendentry{$\sigma_{k}$, linear}
    \addplot[smooth, domain=0.1:1000, dotted,red] {0.6675*((0.202-0.000028*x)/(0.09-0.000025*x))^(0.5)};
           \addlegendentry{$\sigma_{k}$, non-linear}  
        
    \end{semilogxaxis}
    \end{tikzpicture}
from the graphs we see that the two relations diverge for a stress which is about $10^{2}$ times the brittle fracture stress $\sigma_{f}=100\mbox{ N/mm}^{2}$: for a stress which is below this brittle fracture stress (for $\log(\sigma/\sigma_{f})<10^{0}$) the non-linear and the linear relations give the same results.

\subsubsection{Cubic crystals}

For Cubic crystals of the classes $432, \bar{4}3m$ and $m3m$, (the higher-symmetry classes) the piezo-optic tensor can be represented in terms of three moduli:
\begin{equation}\label{Emmecubico}
\Mel=M_{1}\frac{1}{3}\Ib\otimes\Ib+M_{2}\sum_{k=1}^{3}(\Wb_{k}\otimes\Wb_{k}-\frac{1}{3}\Ib\otimes\Ib)+M_{3}\sum_{j=4}^{6}\Wb_{j}\otimes\Wb_{j}
\end{equation}
where
\begin{equation}
M_{1}=\Mel_{1111}+2\Mel_{1122}\,,\quad M_{2}=\Mel_{1111}-\Mel_{1122}\,,\quad M_{3}=\Mel_{1212}\,;
\end{equation}
from (\ref{Emmecubico}) it follows that
\begin{eqnarray}\label{Emmecubico2}
\Mel[\Tb]&=&M_{1}\sigma_{m}\Ib+M_{2}(\hat{T}_{11}\Wb_{1}+\hat{T}_{22}\Wb_{1}+\hat{T}_{33}\Wb_{3})\\
&+&M_{3}(T_{23}\Wb_{4}+T_{13}\Wb_{5}+T_{12}\Wb_{6})\,.\nonumber
\end{eqnarray}

For the lower-symmetry cubic classes $23\,,m3$ in place of (\ref{Emmecubico2}) we have instead
\begin{eqnarray}\label{Emmecubico3}
\Mel[\Tb]&=&M_{1}(\sigma_{m}+\hat{T}_{11})\Wb_{1}+M_{2}(\sigma_{m}+\hat{T}_{22})\Wb_{1}+M_{3}(\sigma_{m}+\hat{T}_{33})\Wb_{3}\nonumber\\
&+&\Mel_{1212}(T_{23}\Wb_{4}+T_{13}\Wb_{5}+T_{12}\Wb_{6})\,,\
\end{eqnarray}
where
\begin{eqnarray}
&M_{1}=\Mel_{1111}+2\Mel_{1122}+\Mel_{1133}\,,\quad M_{2}=\Mel_{1111}+\Mel_{2211}+\Mel_{1122}\,,\\
&M_{3}=\Mel_{1111}+\Mel_{3311}+\Mel_{2211}\,.\nonumber
\end{eqnarray}

If we use (\ref{Emmecubico2}) and(\ref{Emmecubico3}) into the linearized relation  (\ref{nkiso}), then it appears that in absence of shear stress $T_{ij}\,,i\neq j$, the frame $\{\eb_{k}\}\,,k=1,2,3,$ is still a principal frame for $\Bb(\Tb)$ and the principal refractive index still are given by (\ref{linearisotropic}): moreover whenever at least two shear stress are zero we are still able to write in an explicit form the principal refractive index, since we know one of the principal directions.

\subsection{Optically Uniaxial crystals}

For the optically uniaxial crystals,in force of the considerations we did about isotropic materials, we shall deal only with the linearized relation (\ref{linearizzatouniax})$_{2}$; by using the representation (D31) from the Dataset [DS], which gives the tensor $\Mel[\Tb]$ in terms of the six components $N_{K}=N_{K}(\Mel\,,\sigma_{m}\,,\dev\Tb)$ into (\ref{linearizzatouniax})$_{2}$ we arrive at:
\begin{eqnarray}\label{enneexampleuni}
\Nel[\Tb]&=&-\frac{n_{o}^{3}}{2}\Bigg(N_{1}\Wb_{1}+N_{2}\Wb_{2}+N_{6}\Wb_{6}\nonumber\\
&+&(1+F_{1}(\xi)+F_{2}(\xi))N_{3}\Wb_{3}\\
&+&(1+F_{1}(\xi))(N_{4}\Wb_{4}+N_{5}\Wb_{5})\Bigg)\,,\nonumber
\end{eqnarray}
a relation which holds true for any optically uniaxial crystal. Clearly $\Nel[\Tb]$ has not, for a generic $\Tb$, the same eigenvectors of $\Bb_{o}$ and we are left with the problem to solve the eigenvalues problem for the tensor
\begin{equation}
\Nb(\Tb)=n_{o}(\Ib-\Wb_{3})+n_{e}\Wb_{3}+\Nel[\Tb]\,;
\end{equation}
however from (\ref{enneexampleuni}) we can obtain the restrictions on the stress in order that $\Nel[\Tb]$ and $\Bb_{o}$ have at least a common eigenvector. Trivially this can be obtained if two between the components $N_{4}\,,N_{5}$ and $N_{6}$ vanishes. 

For the Trigonal classes $3\,,\bar{3}$ the request that two of these components be zero requires $\hat{T}_{11}=\hat{T}_{22}$ and $T_{ij}=0$, $i\neq j$ which implies that also the third constant vanishes: accordingly $\{\eb_{k}\}$ is a base of eigenvectors for $\Nb(\Tb)$ and hence for $\Bb(\Tb)$. For the classes $32, 3m,\bar{3}m$ when $T_{12}=T_{23}=0$ we have $N_{5}=N_{6}=0$ and $\eb_{1}$ is an eigenvector for $\Bb(\Tb)$.

For the Tetragonal and Hexagonal lower-symmetry classes the condition $T_{13}=T_{23}=0$ makes both $N_{4}=N_{5}=0$ and the symmetry direction $\eb_{3}$ is an eigenvector for $\Bb(\Tb)$, whereas for the high-symmetry classes it is sufficient that two shear stress vanish.

If we consider for instance the Tetragonal $4/m$ lead-tungstate PbWO$_{4}$ (PWO) with $n_{o}=2.270$ and $n_{e}=2.186$ at $\lambda=525$ nm, \cite{BA97} then $\xi=0.962$ with
\begin{equation}
1+F_{1}+F_{2}\approx 1+F_{1}=0.925\,;
\end{equation}
the six components of the tensor $\Nb(\Tb)$ then are given by:
\begin{eqnarray}
N_{11}&=&n_{o}-\frac{n_{o}^{3}}{2}(\sigma_{m}M_{1}+\Mel_{11}\hat{T}_{11}+\Mel_{12}\hat{T}_{22}+\Mel_{13}\hat{T}_{33}+\Mel_{16}T_{12})\nonumber\\
N_{22}&=&n_{o}-\frac{n_{o}^{3}}{2}(\sigma_{m}M_{1}+\Mel_{12}\hat{T}_{11}+\Mel_{11}\hat{T}_{22}+\Mel_{13}\hat{T}_{33}-\Mel_{16}T_{12})\nonumber\\
N_{33}&=&n_{e}-0.925\,\frac{n_{o}^{3}}{2}(\sigma_{m}M_{3}+(\Mel_{33}-\Mel_{31})\hat{T}_{33})\\
N_{12}&=&\Mel_{66}T_{12}\nonumber\\
N_{13}&=&\nonumber0.925\,\Mel_{44}T_{13}\\
N_{23}&=&\nonumber0.925\,\Mel_{44}T_{23}\,.
\end{eqnarray}
The eigenvalues of $\Nb(\Tb)$ can be obtained either in exact form whenever at least two shear stress vanishes or, provided $\|\Bb_{o}\|>>\|\Mel[\Tb]\|$, with approximated methods like the one proposed \emph{e.g.\/} into \cite{PT83}.

\subsection{Optically Biaxial crystals}

Also in this case we deal with the linearized relation (\ref{doublebiaxial}) only, and by a simple calculation with the representation (D31) from the Dataset [DS] we arrive at:
\begin{eqnarray}\label{enneexamplebia}
\Nel[\Tb]&=&n_{3}^{3}\Bigg((G_{3}(\xi_{1}\,,\xi_{2})+G_{4}(\xi_{1}\,,\xi_{2})+G_{6}(\xi_{1}\,,\xi_{2}))N_{1}\Wb_{1}\nonumber\\
&+&(G_{3}(\xi_{1}\,,\xi_{2})+G_{5}(\xi_{1}\,,\xi_{2})+G_{7}(\xi_{1}\,,\xi_{2}))N_{2}\Wb_{2}\nonumber\\
&+&G_{3}(\xi_{1}\,,\xi_{2})N_{3}\Wb_{3}+(G_{3}(\xi_{1}\,,\xi_{2})+G_{5}(\xi_{1}\,,\xi_{2}))N_{4}\Wb_{4}\\
&+&(G_{3}(\xi_{1}\,,\xi_{2})+G_{4}(\xi_{1}\,,\xi_{2}))N_{5}\Wb_{5})\nonumber\\
&+&(G_{3}(\xi_{1}\,,\xi_{2})+G_{4}(\xi_{1}\,,\xi_{2})+G_{5}(\xi_{1}\,,\xi_{2})+G_{8}(\xi_{1}\,,\xi_{2}))N_{6}\Wb_{6}\Bigg)\,.\nonumber
\end{eqnarray}

As in the case of uniaxial crystals, in order that $\Bb(\Tb)$ and $\Bb_{o}$ have at least a common eigenvector we need that two components between  $N_{4}\,,N_{5}$ and $N_{6}$ must vanish. In monoclinic crystals, from the relations for $N_{K}$ provided in [DM] this means that we may have $T_{13}=T_{23}=0$ with the monoclinic $b-$axis $\eb_{3}$ as the common eigenvector. For Orthorhombic crystals instead it suffices that two of the shear stress must vanish. There is no such a possibility for Triclinic crystals instead.

We consider as an example the cerium-doped {L}u$_{x}${Y}$_{2-x}${S}iO$_{5}$ (LYSO) which is monoclinic, class $2/m$ with $n_{1}=1.8313$, $n_{2}=1.8524$ and $n_{3}=1.8277$ at $\lambda=409$ nm \cite{PB12}: in this case $\xi_{1}=1.002$ and $\xi_{2}=1.013$, with:
\[
G_{3}=-0.2304\,,\quad G_{4}=-0.0009\,,\quad G_{5}=-0.0056\,,
\]
\[
G_{6}=-0.0074\,,\quad G_{7}=-0.0010\,,\quad G_{8}=3\cdot 10^{-5}\,,
\]
and the six components of $\Nb(\Tb)$:
\begin{eqnarray}
N_{11}&=&n_{1}-0.238\,\frac{n_{3}^{2}}{2}(\sigma_{m}M_{1}+\Mel_{11}\hat{T}_{11}+\Mel_{12}\hat{T}_{22}+\Mel_{13}\hat{T}_{33}+\Mel_{16}T_{12})\nonumber\\
N_{22}&=&n_{2}-0.237\,\frac{n_{3}^{2}}{2}(\sigma_{m}M_{2}+\Mel_{21}\hat{T}_{11}+\Mel_{22}\hat{T}_{22}+\Mel_{23}\hat{T}_{33}+\Mel_{26}T_{12})\nonumber\\
N_{33}&=&n_{3}-0.230\,\frac{n_{3}^{2}}{2}(\sigma_{m}M_{3}+\Mel_{31}\hat{T}_{11}+\Mel_{32}\hat{T}_{22}+\Mel_{33}\hat{T}_{33}+\Mel_{36}T_{12})\nonumber\\
N_{12}&=&-0.236\,\frac{n_{3}^{2}}{2}(\Mel_{44}T_{23}+\Mel_{45}T_{13})\nonumber\\
N_{13}&=&-0.231\,\frac{n_{3}^{2}}{2}(\Mel_{54}T_{23}+\Mel_{55}T_{13})\nonumber\\
N_{23}&=&-0.236\,\frac{n_{3}^{2}}{2}(\Mel_{16}\hat{T}_{11}+\Mel_{26}\hat{T}_{22}+\Mel_{36}\hat{T}_{33}+\Mel_{66}T_{12})\,;\nonumber
\end{eqnarray}
here for $T_{13}=T_{23}=0$ the eigenvalues can be obtained in explicit form, whereas in the other cases we need an approximate method to find the eigenvalues like the one proposed in \cite{PT83}.

\section{Conclusions}

We first obtained the exact expression for the permittivity tensor and its square root for optically isotropic, uniaxial and biaxial crystals, by applying a result obtained into \cite{HC84}: the principal components of the permittivity square root are the principal refractive index. 

Then we get the linearized relations for both the permittivity tensor and its square root, to within higher-order terms in the stress tensor: these relations holds for any crystallographic symmetry and any stress tensor. By the means of an example concerning glass, which is optically anisotropic, we show that the linearized and the exact relations coincides for stress which two order bigger than the brittle fracture stress. 

We finish by writing the components of the square root of the permittivity tensor for optically uniaxial and biaxial crystals and by showing the restriction on the stress which allow for an explicit evaluation of the principal refraction index, the other cases being dealt with one of the approximate methods which can be found in the literature.

We think that these relations generalize and simplify those presented elsewhere for special cases of stress and crystallographic symmetries.

\section*{Acknowledgments}
The research leading to these results is within the scope of CERN R\&D Experiment 18 "Crystal Clear Collaboration" and the PANDA Collaboration at GSI-Darmstadt. 

\section*{Dataset}
In order to make the paper more concise and readable, many calculations and explicit expressions are presented and collected into this paper Datased [DS] which is available on Mendeley, DOI: 10.17632/3yz353c8ms.2.

\end{document}